\documentclass[epj,nopacs]{svjour}
\usepackage{graphicx}
\usepackage{amsmath,amssymb,color}
\usepackage{bm}
\begin{document}
\title{Three-$\alpha$ configurations of the second $J^\pi=2^+$ state in $^{12}$C}
\author{
  H. Moriya\inst{1} \thanks{\emph{e-mail:} moriya@nucl.sci.hokudai.ac.jp} \and 
  W. Horiuchi\inst{2,3,4,1} \thanks{\emph{e-mail:} whoriuchi@omu.ac.jp} \and 
  J. Casal\inst{5} \and 
  L. Fortunato\inst{6,7}
}
\institute{
Department of Physics, Hokkaido University, Sapporo 060-0810, Japan \and 
Department of Physics, Osaka Metropolitan University, Osaka, 558-8585, Japan \and
Nambu Yoichiro Institute of Theoretical and Experimental Physics (NITEP), Osaka Metropolitan University, Osaka 558-8585, Japan \and
RIKEN Nishina Center, Wako 351-0198, Japan \and 
Departamento de Física At\'omica, 
Molecular y Nuclear, Facultad de Física, Universidad de Sevilla,Apartado 1065, E-41080 Sevilla, Spain \and 
Dipartimento di Fisica e Astronomia “G.Galilei”, Universit degli Studi di Padova, via Marzolo 8, Padova I-35131, Italy \and 
INFN-Sezione di Padova, via Marzolo 8, Padova I-35131, Italy
}
\date{Received: date / Revised version: date}
\abstract{
  We investigate geometric configurations of $\alpha$ ($^4$He nucleus) clusters in
  the second $J^\pi=2^+$ state of $^{12}$C, 
  which has been discussed as a rotational band member of
  the second $0^+$ state, the Hoyle state.
  The ground and excited $0^+$ and $2^+$ states
  are described by a three-$\alpha$ cluster model.
  The three-body Schr\"odinger equation with orthogonality conditions
  is accurately solved by the stochastic variational method with correlated Gaussian basis functions.
  To analyse the structure of these resonant states
  in a convenient form, we introduce a confining potential.
  The two-body density distributions together with the spectroscopic information
  clarify the structure of these states.
  We find that main configurations of both the second $0^+$ and $2^+$ states are
  acute-angled triangle shapes originating from the $^8$Be($0^+$)$+\alpha$ configuration.
  However, the $^{8}{\rm Be}+\alpha$ components in the second $2^+$ state
  become approximately 2/3 because the $^8$Be subsystem is hard to excite,
  indicating that the state is not 
  an ideal rigid rotational band member of the Hoyle state.
} 
\maketitle
\section{Introduction}
An $\alpha$ ($^4$He nucleus) cluster is one of the most fundamental
ingredients for understanding the structure of nuclei.
The first excited $J^\pi=0^+$ state of $^{12}$C, the so-called Hoyle state, 
is believed to play a crucial role
in generating the $^{12}$C element in the universe~\cite{Hoyle54}.
For more than half a century, the Hoyle state has been studied
by various theoretical models. 
As the state has a significant amount of the
$^8$Be($0^+$)+$\alpha$ configurations~\cite{Horiuchi74,Horiuchi75,Uegaki77,Uegaki78,Uegaki79},
the Hoyle state decays dominantly via sequential decay process
$^8{\rm Be}(0^+)\alpha\to 3 \alpha$~\cite{Ishikawa14}.
On the other hand, Ref.~\cite{Tohsaki01} claimed
that the Hoyle state has the $\alpha$-condensate-like character,
where three $\alpha$ bosons occupy the same $S$ orbit.
The structure of the Hoyle state has also been discussed in terms of
geometric configurations of three-$\alpha$ particles based on the 
algebraic cluster model (ACM)~\cite{Bijker02,Fortunato19,Vitturi20}.
Fully microscopic calculations predicted a significant
amount of $\alpha$ cluster configurations
in the Hoyle state~\cite{Chernykh07,Kanada07}.
Prominent three-$\alpha$ cluster structure configurations
were confirmed in density functional theory~\cite{Ebran13,Ebran20}
and very recently in the Monte Carlo Shell Model approach~\cite{Otuka22}
The evidence of the three-$\alpha$ cluster structure
can also be seen in its density profile of the ground state~\cite{Horiuchi23}.

The search for other excited cluster states with some 
analogy to the Hoyle states has attracted interest.
The structure of the second $J^\pi=2^+$ state
is controversial as it can be a candidate of a rotational excited state 
of the Hoyle state forming the ``Hoyle band''~\cite{Freer11}.
Experimentally, the existence of the $2_2^+$ state was confirmed~\cite{Itoh04,Freer09,Itoh11,Zimmerman13}
at 2.59(6)MeV above the three-$\alpha$ threshold with the decay width of ~1.01(15)MeV~\cite{N12ND}.
The idea of the Hoyle band has attracted attention.
Ref.~\cite{Smith20} deduced
a limit for the direct decay branching ratio of the Hoyle state
under the assumption that the intrinsic structure
of $0_2^+$ and $2_2^+$ are the same.
Theoretically,  the $2_2^+$ state has only been
recognized as having dominant  $^{8}\mathrm{Be}(0^+) + \alpha$ configurations,
in which its intrinsic structure is a weakly-coupled $^{8}$Be
plus an $\alpha$ particle 
with the angular momentum of 2~\cite{Uegaki77,Uegaki78,Uegaki79,Kanada07}.
In analogy to the Hoyle state, the $\alpha$-mean field character
in the $2_2^+$ state
can be considered, in which one $\alpha$ particle
is excited to the $D$ orbit~\cite{Funaki05,Yamada05}
but Ref.~\cite{Funaki15} argued that the $2_2^+$ state is not 
a simple rigid rotational excited state based on the analysis of 
the energy levels
obtained by the microscopic three-$\alpha$ cluster model.
In the context of the ACM, the $2_2^+$ state is interpreted 
as the rigid rotational excited state of the Hoyle state in which 
three $\alpha$ particles geometrically form an equilateral triangle 
and vibrate with the $\mathcal{D}_{3h}$ symmetry~\cite{Bijker20}.
To confirm whether this state belongs to the Hoyle state, 
a certain degree of 
similarity in the intrinsic structure should be observed.
This motivates us to conduct a detailed study to clarify 
the extent of similarity between
the structure of the second $0^+$ and $2^+$ states.

To settle this argument, in this paper,
we study geometric configurations of three-$\alpha$ particles
in the second $2^+$ state and compare its structure with the second
$0^+$ Hoyle state using accurate three-$\alpha$ wave functions.
$^{8}{\rm Be}+\alpha$ components are analysed to clarify
the origin of these configurations.

In this paper,
the four physical states, 
$J^\pi=0_1^+$, $0_2^+$, $2_1^+$ and $2_2^+$
of $^{12}$C are studied within the three-$\alpha$ cluster model.
In the next section, we explain our approach.
Fully converged solutions are obtained by correlated Gaussian expansion
with the stochastic variational method.
They are briefly explained in Secs.~\ref{method1.sec} and \ref{method2.sec}.
Geometric configurations of the $\alpha$ particles
are visualized by calculating
two-body density distributions as well as other physical quantities.
To evaluate these physical quantities of the state with
rather a wide decay width such as the second $2^+$ state,
we introduce a confining potential.
The details are given in Sec.~\ref{method3.sec}.
In Sec.~\ref{results.sec}, we show the numerical results and analysis.
Finally, we draw a conclusion about
the structure of the $2_2^+$ state
in Sec.~\ref{conclusions.sec}.

\section{Method}
\label{method.sec}

\subsection{Three-$\alpha$ cluster model}
\label{method1.sec}

In this paper, the wave functions of $^{12}$C are described
as a three-$\alpha$ system. The three-$\alpha$ Hamiltonian reads 
\begin{equation}
  H=\sum_{i=1}^{3} T_{i}-T_{\mathrm{cm}} + \sum_{i>j=1}^{3} (V_{2\alpha}^{ij}+V_{\rm Coul.}^
  {ij})+ V_{3\alpha},
\end{equation}
where $T_{i}$ is the kinetic energy
of the $i$th $\alpha$ particle. The kinetic 
energy of the center-of-mass motion $T_{\mathrm{cm}}$ is subtracted.
The mass parameter in the kinetic energy terms
and the elementary charge in the Coulomb potential
$(V_{\rm Coul.})$ are taken as
$\hbar^2/m_{\alpha}=10.654$ MeVfm$^2$ and 
$e^2=1.440$ MeVfm, respectively.
Two-$\alpha$ interaction $V_{2\alpha}$
is taken as the same used in Ref.~\cite{Fukatsu92},
which is derived by a folding procedure using
an effective nucleon-nucleon interaction.
We employ the three-$\alpha$ interaction $V_{3\alpha}$ depending
on the total angular momentum $J^{\pi}$ reproducing
the energies of the $0_1^+$ and $2_1^+$ states
as was used in Ref.~\cite{Ohtubo13}.
Here we adopt the orthogonality condition model~\cite{Saito68,Saito69,Saito77}.
To impose the orthogonality condition to the Pauli forbidden states (f.s.),
we introduce in the Hamiltonian the following pseudopotential~\cite{Kukulin78}:
\begin{align}
  V_{\mathrm{P}}
  &=\gamma \sum_{i>j=1}^{3} \sum_{nlm \in \mathrm{f.s.}} \vert\phi_{nlm}(ij)\rangle \langle \phi_{nlm}(ij) \vert.
\end{align}
The summation of $nlm$ runs over all the f.s., i.e., $0S$, $1S$, and $0D$ states.
We adopt the harmonic oscillator wave functions
for $\phi_{nlm}$ with the size 
parameter $\nu=0.2575$ fm$^{-2}$~\cite{Fukatsu92}
reproducing the size of the $\alpha$ particle.
Taking $\gamma$ large enough, we exclude the Pauli
forbidden states variationally from numerical calculations.
In this paper, we take $\gamma=10^5$ MeV.
The f.s. components of the resulting wave functions are
found to be in the order of $10^{-5}$.

\subsection{Correlated Gaussian expansion}
\label{method2.sec}

Denoting the $i$th single $\alpha$ particle coordinate vector
by $\bm{r}_i$ $(i=1,2,3)$, we define
a set of Jacobi coordinates  
$\bm{x}_1=\bm{r}_2-\bm{r}_1$ and $\bm{x}_2=\bm{r}_3-(\bm{r}_1+\bm{r}_2)/2$,
excluding the center-of-mass
coordinate $\bm{x}_3=(\bm{r}_1+\bm{r}_2+\bm{r}_3)/3$, which
are denoted as $\tilde{\bm{x}}=(\bm{x}_1,\bm{x}_2)$, where
a tilde stands for the transpose of a matrix.
The $k$th state of
the three-$\alpha$ wave function $\Psi_{JM}^{(k)} (\bm{x})$
with the total angular momentum $J$ and its projection $M$
is expressed in a superposition of fully symmetrized correlated Gaussian
basis functions $G$~\cite{Varga95,SVMbook},
\begin{align}
    \Psi_{JM}^{(k)}&=\sum_{i=1}^{K} C_i^{(k)} G(A_i, u_i, \bm{x}),  \\
    G(A_i,u_i,\bm{x}) &= \mathcal{S} \exp \left(-\frac{1}{2}\tilde{\bm{x}} A_i \bm{x}\right)\mathcal{Y}_{JM}(\tilde{u}_i \bm{x}),
\end{align}
where $\mathcal{S}$ is the symmetrizer which
makes basis functions symmetrized under all
particle-exchange, ensuring bosonic properties
of $\alpha$ particles.
A variational parameter $A_i$ is a 2 by 2
positive definite symmetric matrix, and
$\tilde{\bm{x}}A\bm{x}$ is a short-hand notation of
$\sum_{i,j=1}^2A_{ij}\bm{x}_i \cdot \bm{x}_j$.
The angular part of the wave function is described by using 
the global vector $\tilde{u}\bm{x}=\sum_{j=1}^{2}u_{j} \bm{x}_j$
with $\tilde{u}=(u_{1},u_{2})$ and $u_2^2=1-u_1^2$~\cite{SVMbook,Suzuki98}.
A set of linear coefficients ${C_i^{(k)}}$
is determined by solving the generalized 
eigenvalue problem,
\begin{equation}
    \sum_{j=1}^{K}H_{ij} C_{j}^{(k)}=E^{(k)} \sum_{j=1}^{K}B_{ij} C_{j}^{(k)} \quad (i=1,\ldots,K),
\end{equation}
where the matrix elements $H_{ij}$ and $B_{ij}$ are defined as 
\begin{align}
    &H_{ij}=\langle G(A_i,u_i,\bm{x}) \vert H \vert G(A_{j},u_j,\bm{x})\rangle \\
    &B_{ij}=\langle G(A_i,u_i,\bm{x}) \vert G(A_{j},u_j,\bm{x}) \rangle.
\end{align}
The variational parameters $A_i$ and $u_i$ are determined
by the stochastic variational method~\cite{Varga95,SVMbook}.
For more details of the optimization procedure,
the reader is referred to Refs.\cite{Phyu20,Moriya21}.

\subsection{Confining potential}
\label{method3.sec}

In this paper, we treat resonant $0_2^+$ and $2_2^+$ states as a bound state.
This is the so-called bound-state approximation and works well
for a state with a narrow decay width such as the $0_2^+$
state (Expt.: $\Gamma=8.5\times 10^{-3}$ MeV~\cite{Ajzenberg90}), 
while for the $2_2^+$ state it is hard to obtain the physical state with a simple basis expansion~\cite{Funaki06}
as it has somewhat a large decay width 
(Expt.: $\Gamma=1.01(15)$ MeV~\cite{Itoh11}).
To estimate the resonant energy,
the analytical continuation in the coupling constant~\cite{ACCC}
is useful but does not provide us with the wave function.
Nevertheless, a square-integrable wave function of
a resonant state
is useful to analyse its structure.
A confining potential (CP) method~\cite{Mitroy08,Mitroy13} is suitable 
for this purpose,
as we can treat a resonance state as a bound state inside of the CP.
To get a physical resonant state in the bound-state approximation,
we introduce the CP in the following parabolic
form~\cite{Mitroy08} as
\begin{equation}
    V_{\mathrm{CP}}=\sum_{i=1}^{3}\lambda \Theta(\vert\bm{r}_{i}-\bm{x}_3\vert-R_0) (\vert\bm{r}_{i}-\bm{x}_3\vert-R_0)^2,
\end{equation}
where $\Theta(r)$ is the Heaviside step function,
\begin{equation}
    \Theta(x) = \begin{cases}
        1 & (x > 0) \\
        0 & (x < 0)
       \end{cases}.
\end{equation}
The strength $\lambda$ and range $R_0$ parameters
of the CP are real numbers
and have to be taken appropriately.
Here we investigate the stability of the energies
as well as the root-mean-square (rms) radii of $\alpha$ particles
$R_{\rm rms}= \sqrt{\langle\Psi_{JM}
  \vert(\bm{r}_1-\bm{x}_3)^2\vert\Psi_{JM}}\rangle$ 
of the $0_1^+$, $0_2^+$, $2_1^+$, and $2_2^+$ states against changes of $\lambda$ and $R_0$.

\begin{figure}
\resizebox{0.5\textwidth}{!}{\includegraphics[bb=0.000000 0.000000 274.475000 216.985625]{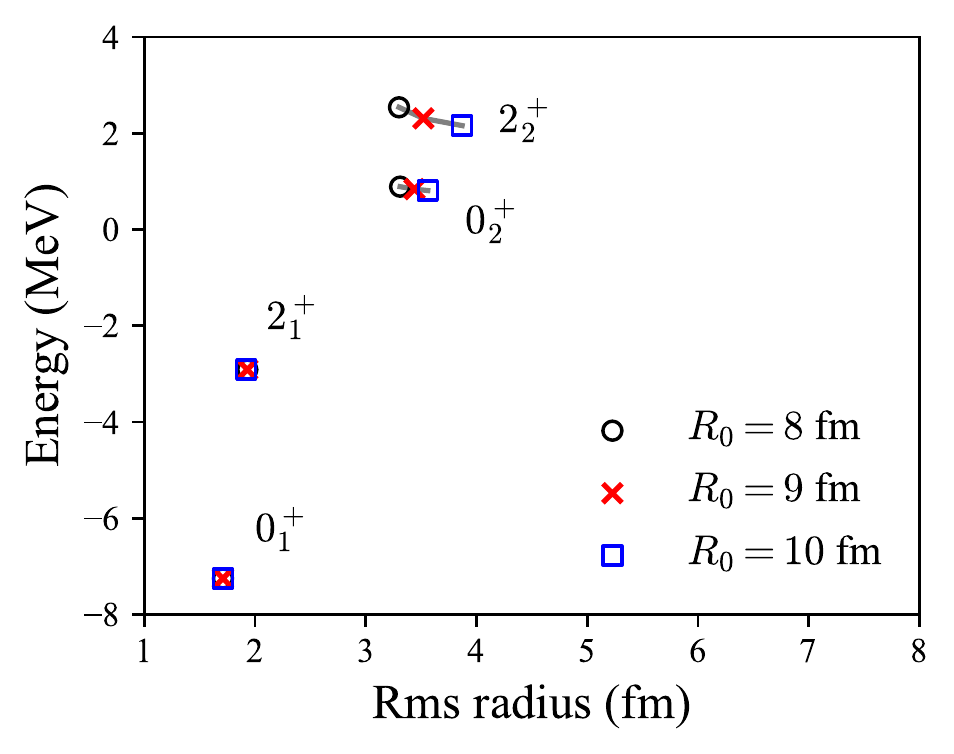}}
\caption{$R_0$ dependence in the CP. Energies and rms radii of 
the $0_1^+$, $0_2^+$, $2_1^+$, and $2_2^+$ states
with $R_0=8, 9$ and 10~fm are plotted. 
The strength of the confining potential $\lambda$
is set to be 100 MeV/fm$^2$. See text for details.}
\label{fig:confine}
\end{figure}

Figure~\ref{fig:confine} shows
the energies and rms radii 
of the $0_1^+$, $0_2^+$, $2_1^+$ and $2_2^+$ states
with different $R_0$.
The strength of the confining potential is set to be
$\lambda=100$ MeV/fm$^2$.
Since the $R_0$ value is taken large enough,
the energies and the rms radii of the 
bound states, the $0_1^+$ and $2_1^+$ states,
do not depend much
on these parameters.
Even for the resonant $0_2^+$  and $2_2^+$ states,
we find that the fluctuations of the energies
are small about 0.1~MeV and 0.6~MeV, respectively,
in the range of $R_0=8$--10 fm.
This is reasonable considering the facts that the $0_2^+$ state has
a quite small decay width
and the $2_2^+$ state has a larger decay width.
The magnitude of the radius fluctuation
against the changes of $R_0$ is about
$\approx 0.3$ fm for the $0_2^+$ state
and $\approx 0.5$ fm for the $2^+_2$ state.
We also made the same analysis by strengthening the
strength $\lambda$ by 10 times
and a similar plot was obtained.
Hereafter, we use the results with
$R_0=9$~fm, $\lambda=100$~MeV/fm$^2$.

Table~\ref{tab:erd} lists the calculated energies and rms radii
of $\alpha$ particles.
These energy values can be compared
with the real parts of the complex energies
obtained by the complex scaling method (CSM)~\cite{Ohtubo13}.
The energies are 0.75 and 2.24 MeV for the
$0^+_2$ and $2_2^+$ states, respectively,
which are in good agreement with our results.
Finally, we obtain the rms radii of the $0_2^+$ and $2_2^+$ states
using these obtained wave functions. They are found to be similar
and significantly large compared to the $0_1^+$ and $2_1^+$ states.
For the sake of convenience, we also list 
the charge radii evaluated
by $R_{\rm ch}=\sqrt{r_{\alpha}^2+R_{\rm rms}^2}$,
where $r_{\alpha}$ is the charge radius of an $\alpha$ particle,
1.6755(28) fm~\cite{Angeli13}.
The calculated result for the ground state agrees
with the theoretical result~\cite{Kurokawa07}, showing
reasonable reproduction of the measured charge-radius
data 2.4702(22) fm~\cite{Angeli13}.
A large charge radius of the $0_2^+$ state is also consistent with
that obtained in Ref.~\cite{Kurokawa07}
though its radius was given as a complex number.

\begin{table}
\centering
\caption{Calculated energies measured from the three-$\alpha$ threshold,
rms radii of $\alpha$ particles, and charge radii} of the $0_1^+$, $0_2^+$, $2_1^+$, and $2_2^+$ states.
\label{tab:erd}
\begin{tabular}{cccc}
\hline\noalign{\smallskip}
$J^{\pi}$ & $E$ (MeV) & $R_{\rm rms}$ (fm)  & $R_{\rm ch}$ (fm) \\
\noalign{\smallskip}\hline\noalign{\smallskip}
$0_1^+$   & $-$7.25              & 1.71   &   2.39   \\
$0_2^+$   & 0.84                 & 3.44   &   3.83   \\
$2_1^+$   & $-$2.92              & 1.93   &   2.56   \\
$2_2^+$   & 2.32                 & 3.50   &   3.88   \\
\noalign{\smallskip}\hline
\end{tabular}
\end{table}

\section{Results}
\label{results.sec}

\subsection{Three-$\alpha$ configurations: Two-body density}

\begin{figure*}
\centering
\includegraphics[bb=0.000000 0.000000 254.800000 258.610625, width=0.4\linewidth]{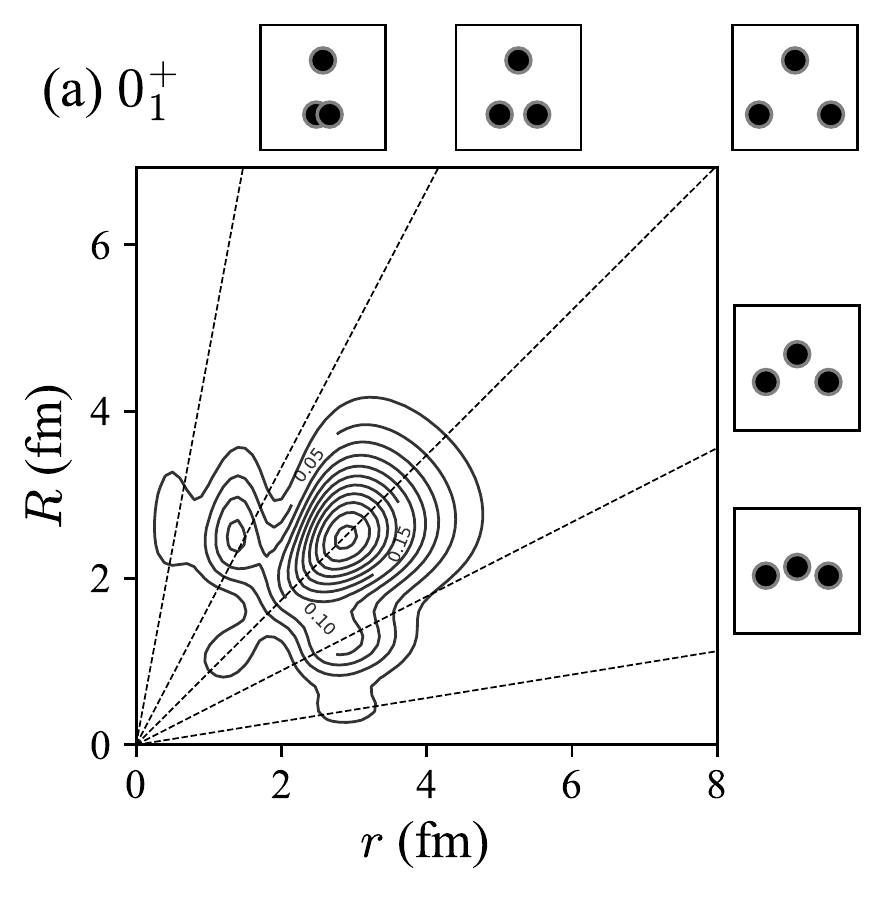}
\includegraphics[bb=0.000000 0.000000 254.800000 258.610625, width=0.4\linewidth]{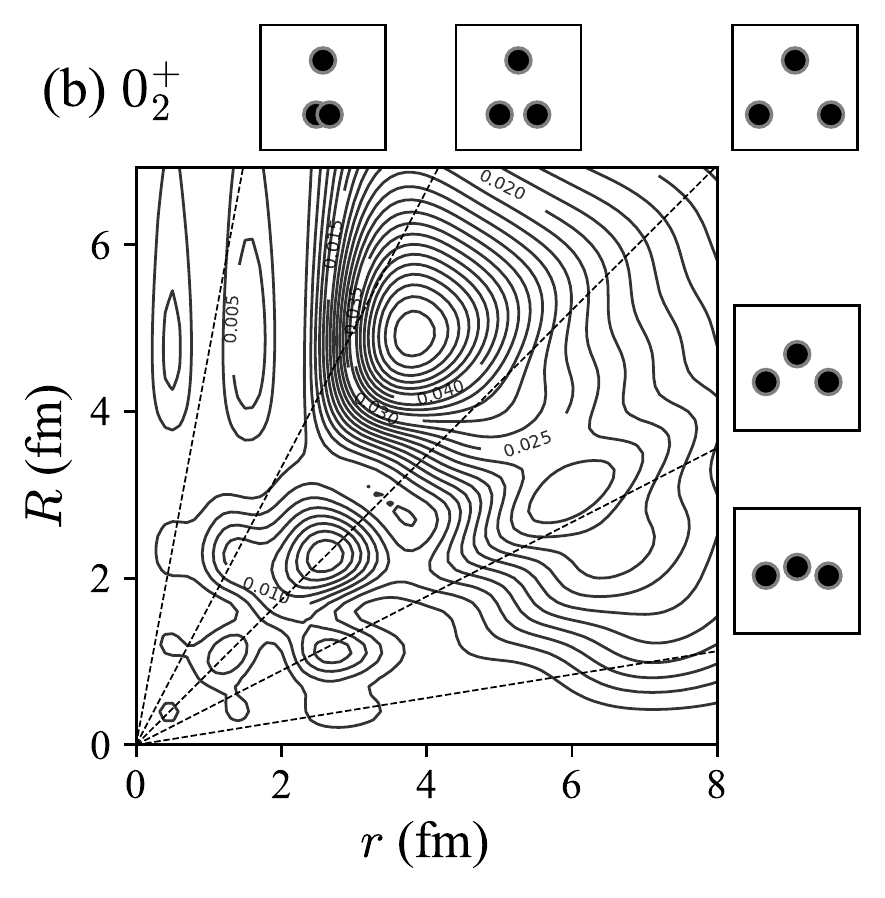}
\includegraphics[bb=0.000000 0.000000 254.800000 258.610625, width=0.4\linewidth]{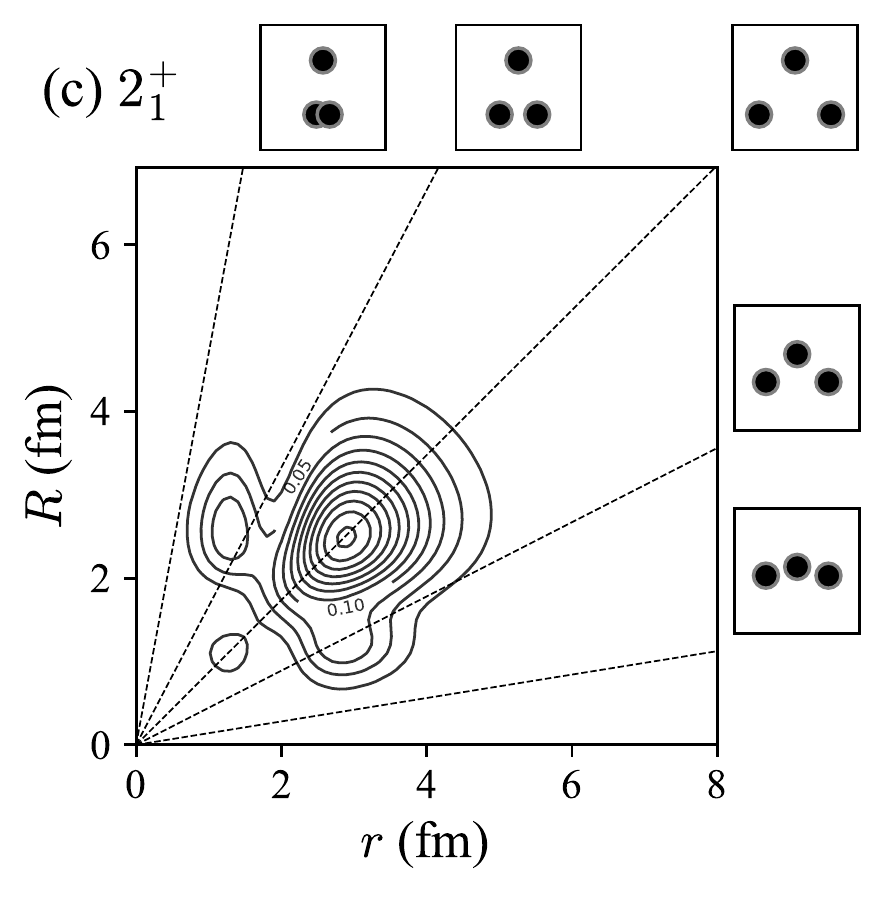}
\includegraphics[bb=0.000000 0.000000 254.800000 258.610625, width=0.4\linewidth]{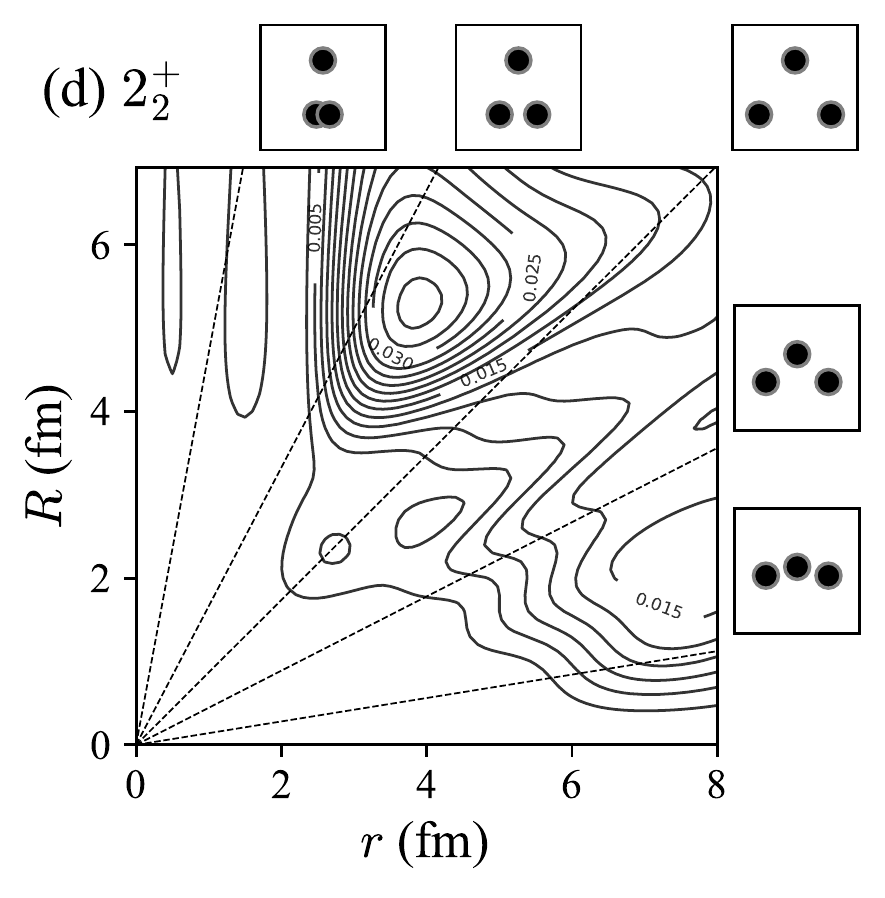}
\caption{\label{fig:TBD}
  Two-body density distributions
  $\rho(r,R)$ of the (a) $J^{\pi}=0_1^+$,
  (b) $0_2^+$, (c) $2_1^+$, and (d) $2_2^+$ states.
  Contour intervals are 0.025 fm$^{-2}$ for $0_1^+$ and $2_1^+$
  and 0.0025 fm$^{-2}$ for $0_2^+$ and $2_2^+$.
  Specific $r/R$ ratios are indicated by dashed lines
  and their geometric configurations are illustrated in small panels,
  e.g., the diagonal dashed line indicates the equilateral triangle configurations.}
\end{figure*}

To discuss the geometric configurations of the three-$\alpha$ systems,
it is intuitive to see the two-body density distributions
with respect to the two relative coordinates, $x_1$ and $x_2$, defined by 
\begin{equation}
  \rho (r,R) = \langle\Psi\vert\delta(\vert\bm{x}_1\vert-r)\delta(\vert\bm{x}_2\vert-R)\vert\Psi\rangle,
\end{equation}
Note that the distribution is normalized as
$\int_0^{\infty}dr\int_0^{\infty}dR\,\rho(r,R)=1$.
Figure~\ref{fig:TBD} plots the two-body density distributions
of the $J^{\pi}=0_1^+$, $0_2^+$, $2_1^+$, and $2_2^+$ states.
For a guide to the eyes, the specific $r/R$ ratios
are indicated by the dashed lines and their corresponding geometric shapes are
depicted by inset figures.
We remark that the two-body density distributions were already discussed 
for the $J^\pi=0^+$ states in detail
by using the shallow potential models~\cite{Ishikawa14,Nguyen13}.
Here we present the results with the OCM.
The preliminary results for the $0^+$ states were already discussed
in Ref.~\cite{Moriya21FB} but we repeat it to remind
the characteristics of the two-body density distributions
and to compare it with the $2^+$ state.

The two-body density distributions of the $0_1^+$ and $2_1^+$ states
have similar peak structures; the most dominant peak is located on the 
equilateral triangle configuration at $r \sim$ 3~fm and
some other peaks come from the nodal behavior of wave function 
due to the orthogonality of the forbidden states.
We see different fine structures when a shallow potential
model is employed. See Ref.~\cite{Moriya21FB} for detailed comparison.

In contrast to the compact ground state, 
the two-body density distribution of the $0_2^+$ state is widely spreading.
The most dominant peak of the $0_2^+$ state distribution is
located at the acute-angled triangle configuration, 
which comes from the $^{8}{\rm Be}(0^+)+\alpha$ structure~\cite{Moriya21FB}.
For the $2_2^+$ state, likely to the $0_2^+$ state, the two-body density distribution
spreads and the most dominant peak is located
at the acute-angled triangle configuration.
However, we find that the amplitude is significantly smaller
than the $0^+_2$ state.
The difference of these peak structures between the $0_2^+$ and $2_2^+$
states implies different intrinsic structure,
which will be discussed in the next subsection.

At a closer look,
we see the small peaks in the internal regions for
the $0^+_2$ state, while they disappear for the $2_2^+$ state.
This peak structure comes from the occupation
of the nodal $S$ orbit but the occupation number in the $2_2^+$ state
is much smaller than that of the $0_2^+$ state~\cite{Yamada05}.
Because the $2_1^+$ state already has a large occupation number of
the $S$ orbit, there is no space to accommodate the nodal $S$ orbit
in the $2_2^+$ state which should be orthogonal to the $2_1^+$ state.

\subsection{Partial-wave and $^{8}$Be components in the three-$\alpha$ wave functions}

In this subsection, we discuss more detailed structure
of  these three-$\alpha$ wave functions.
For this purpose it is convenient to calculate
the partial-wave component and $^{8}$Be spectroscopic factor,
which are respectively defined by
\begin{align}
  P_{l_1l_2}&=\frac{3!}{2!1!}\lvert\left<[Y_{l_1}(\hat{\bm{x}}_1)Y_{l_2}(\hat{\bm{x}}_2)]_{JM}\vert\Psi_{JM}\right>\rvert^2, \label{eq:PW}\\
  S_{l_1l_2}&=\frac{3!}{2!1!}\lvert\left<\phi_{l_1}(x_1)[Y_{l_1}(\hat{\bm{x}}_1)Y_{l_2}(\hat{\bm{x}}_2)]_{JM}\vert\Psi_{JM}\right>\rvert^2,
  \label{eq:SF}
\end{align}
where $\phi_{l}$ is the radial wave functions of $^8\mathrm{Be}$
with the relative angular momentum $l=0, 2$, or 4,
which correspond to physical resonant states
with $J^\pi=0^+, 2^+$ or $4^+$, respectively,
obtained by solving the two-$\alpha$ system using the 
same two-$\alpha$ potential adopted in this paper.
The CP is also applied to evaluate these resonant states,
and hence the obtained wave functions are square-integrable.
The $P_{l_1l_2}$ value is the probability of
finding $(l_1,l_2)$ component in the three-$\alpha$ wave function,
while the $S_{l_1l_2}$ value can be a measure
of the the $^{8}{\rm Be}+\alpha$ clustering.
Note that given $l_1$ and $l_2$, $S_{l_1l_2}$ is a subspace 
of $P_{l_1l_2}$, hence $S_{l_1l_2}\leq P_{l_1l_2}$ always holds.

Table~\ref{tab:sfpw}
lists the $P_{l_1l_2}$ and $S_{l_1l_2}$ values
for the $0^+$ and $2^+$ states.
The $0_1^+$ state has almost equal
$P_{l_1l_2}$ values for $l_1=l_2=0, 2$, and 4, which can be 
explained by reminding that
the state has the SU(3)-like character~\cite{Yamada05}.
The higher partial-wave components is found to be $\approx 1$\%.
The $0^+_1$ wave function has about 50\% of the $^{8}{\rm Be}+\alpha$ component.
The $2^+_1$ state is mainly composed of
$(l_1,l_2)=(2,2)$ and (4,4) components,
$P_{22}$ and $P_{44}$,
reflecting SU(3) character as like the $0_1^+$ state~\cite{Yamada05}
and also contains about half of the $^{8}{\rm Be}+\alpha$ component.
Consequently, the structure of the $2_1^+$ state
can be interpreted as
a rigid rotational excited state of the $0_1^+$
while keeping its geometric shape as was shown in
Fig.~\ref{fig:TBD}. 

On contrary, the $P_{l_1l_2}$ values of $0_2^+$ 
concentrate only on the $l_1=l_2=0$ channel about 80\%, 
which is consistent with the microscopic cluster model
calculations~\cite{Matsumura04,Yamada05}.
This characteristic behavior is often interpreted
as the bosonic condensate state of
the three-$\alpha$ particles~\cite{Tohsaki01,Yamada05}.
This $(l_1,l_2)=(0,0)$ channel
mostly consists of the $^{8}{\rm Be}(0^+)+\alpha$ component
shown in Table~\ref{tab:sfpw}, forming the acute-angled triangle shape
in the two-body density distribution~\cite{Moriya21FB}.

For the $2_2^+$ state, dominant partial-wave components
are the $(l_1,l_2)=(0,2)$ and (2,0) channels.
The $^{8}{\rm Be}(0^+)+\alpha$ component is dominant
in the $(l_1,l_2)=(0,2)$ channel, while
few $^{8}{\rm Be}(2^+)+\alpha$ component
is found in the $(l_1,l_2)=(2,0)$ channel,
which is in contrast to
the $0_2^+$ state mainly consisting
of the $^{8}{\rm Be}+\alpha$ configuration.
This strong suppression can naturally be understood by
considering the fact that the excitation energy
of $^8\mathrm{Be}(2^+)$ is rather high 3.26 MeV
(Expt.: 3.12 MeV~\cite{Tilley04}),
compared to the calculated energy 
spacing between the $0_2^+$ and $2_2^+$ states, $\approx 1.4$ MeV.
This suggests that the $2_2^+$ state
is not a simple rigid rotational excited state
of the $0_2^+$ state but a partially rotational state.
We remark that this interpretation supports
the mean-field-like picture: 
The three $\alpha$ particles occupy the same $S$ state in
the $0_2^+$ state~\cite{Tohsaki01}, while one $S$-state $\alpha$ particle
is excited to the $D$ state in the $2_2^+$ state~\cite{Yamada05}.
Such a $D$-wave excitation is possible
with lower energy than the $^8$Be excitation
if the frequency of the mean-field potential is low enough.

\begin{table*}[hbt]
  \centering
  \caption{\label{tab:sfpw} Partial-wave component 
  and $^8$Be spectroscopic factor of the $J^\pi=0^+$ and $2^+$ states. 
  See text for details.}
  \begin{tabular}{ccccccccc}
    \hline\noalign{\smallskip}
    &\multicolumn{2}{c}{$0_1^{+}$}&\multicolumn{2}{c}{$2_1^+$}&\multicolumn{2}{c}{$0_2^{+}$}&\multicolumn{2}{c}{$2_2^+$}\\ 
    \cline{2-3} \cline{4-5} \cline{6-7} \cline{8-9}\noalign{\smallskip}  
    $(l_1 l_2)$ & $P_{l_1l_2}$ & $S_{l_1l_2}$ & $P_{l_1l_2}$ & $S_{l_1l_2}$  & 
                  $P_{l_1l_2}$ & $S_{l_1l_2}$ & $P_{l_1l_2}$ & $S_{l_1l_2}$  \\ 
    \noalign{\smallskip}\hline\noalign{\smallskip}              
    (00) &0.352&0.193&--&--&0.786&0.668&--&-- \\
    (02) &--&--&0.096&0.058&--&--&0.451&0.419 \\
    Subtotal ($l_1=0$) &0.352&0.193&0.096&0.058&0.786&0.668&0.451&0.419 \\ \hline
    (20) &--&--&0.095&0.054&--&--&0.374&0.021 \\
    (22) &0.351&0.175&0.483&0.268&0.112&0.027&0.044&0.011 \\
    (24) &--&--&0.006&0.003&--&--&0.020&0.007 \\
    Subtotal ($l_1=2$)&0.351&0.175&0.584&0.325&0.112&0.027&0.438&0.039 \\ \hline
    (42) &--&--&0.007&0.003&--&--&0.029&0.007 \\
    (44) &0.285&0.100&0.299&0.114&0.060&0.013&0.017&0.008 \\
    (46) &--&--&$\sim10^{-4}$&$\sim10^{-5}$&--&--&0.006&0.004 \\
    Subtotal ($l_1=4$)&0.285&0.100&0.306&0.117&0.060&0.013&0.052&0.019 \\ \hline
    Total &0.988&0.468&0.986&0.500&0.958&0.708&0.941&0.477 \\
    \noalign{\smallskip}\hline
  \end{tabular}
\end{table*}

\subsection{Spectroscopic amplitude}

To discuss the role of the dominant channels in the
geometric configurations in the $0_2^+$ and $2_2^+$ states,
it is useful to evaluate the $^8$Be spectroscopic amplitude (SA)
\begin{align}
  &\theta_{l_1l_2}(R) = \sqrt{\frac{3!}{2!1!}}\frac{1}{R}\notag\\&\times\langle\phi_{l_1}(x_1)\left[Y_{l_1}(\hat{\bm{x}}_1)Y_{l_2}(\hat{\bm{x}}_2)\right]_{JM}
    \delta (\lvert\bm{x}_2\rvert-R) \vert\Psi_{JM}\rangle.
    \label{eq:SA}
\end{align}
Note that $\int_0^\infty dR\, [R\theta_{l_1l_2}(R)]^2=S_{l_1l_2}$.
For practical calculations, see
Appendix A of Ref.~\cite{Suzuki09},
where an explicit formula of the SA with the correlated
Gaussian basis function was given.

\begin{figure*}
  \begin{center}
  \includegraphics[bb=0.000000 0.000000 281.100000 217.266875, width=0.45\linewidth]{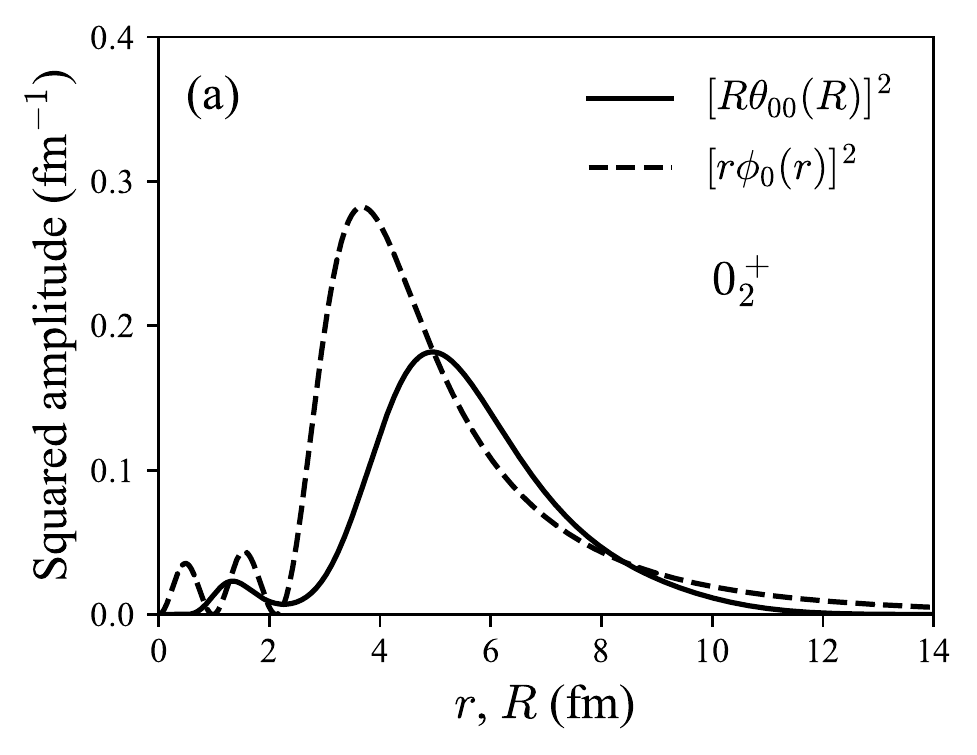}
  \includegraphics[bb=0.000000 0.000000 281.100000 217.266875, width=0.45\linewidth]{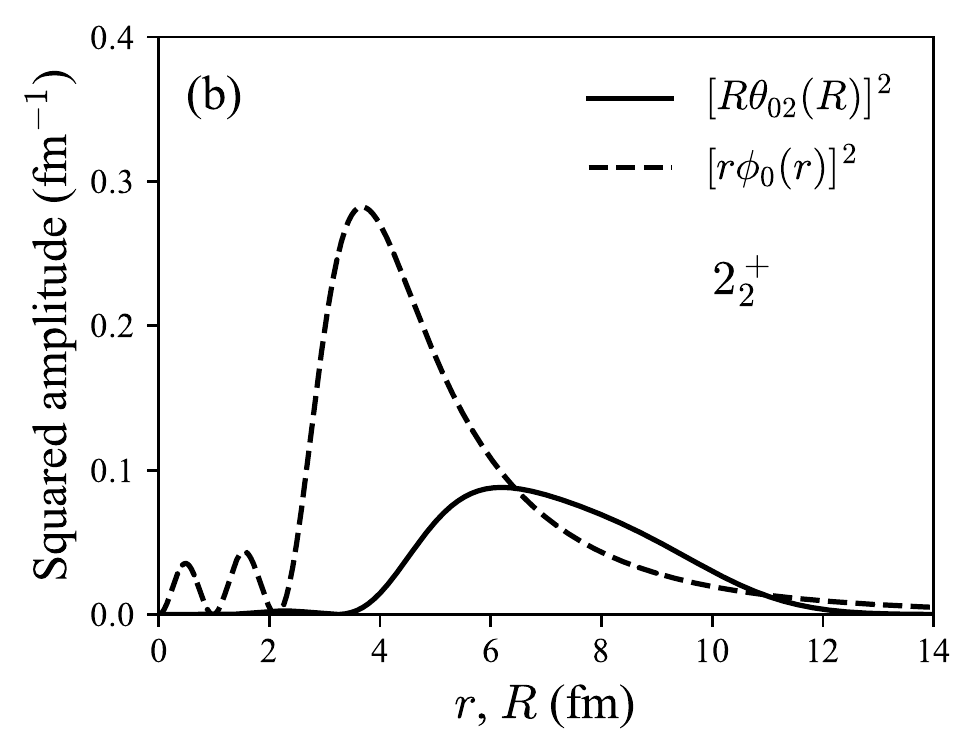}
  \caption{\label{fig:amp}
    Square of $^8$Be spectroscopic amplitudes, 
    $[R\theta_{l_1l_2}(R)]^2$ with (a)
    $(l_1,l_2)=(0,0)$ and (b) $(l_1,l_2)=(0,2)$ for the
    $0_2^+$ and $2_2^+$ states. The square of the
    radial wave function
    of the $^8$Be($0^+$) state $[r\phi_0(r)]^2$ 
    is also compared.
  }
  \end{center}
\end{figure*}

Figure~\ref{fig:amp} shows
the SA with $(l_1,l_2)=(0,0)$
for the $0_2^+$ state and (0,2) for
the $2_2^+$ state, which respectively
correspond to the dominant configurations for each state.
The SA of the $2^+_2$ state is smaller
than that of the $0^+_2$ state reflecting the magnitudes
of the $S_{l_1l_2}$ values.
For the sake of comparison, we also plot
the radial wave function of
$^8$Be($0^+$), $\phi_0(r)$.
The peak position of $r\phi_0(r)$ is located at 3.68 fm,
while the SA has the largest peak at 4.97 fm
for the $0_2^+$ state and 6.20 fm for the $2_2^+$ state.
Though the latter distribution is broad,
these are consistent with the fact that
the highest peak of the two-body density distribution
is located at $(r,R)=(3.9,5.1)$ fm for the $0_2^+$ state
and $(r,R)=(3.9,5.3)$ fm for the $2_2^+$ state,
exhibiting the acute-angled triangle configuration as shown in
Fig.~\ref{fig:TBD}.

We also evaluate the rms radii of 
the SA defined by $D_{l_1l_2}=\sqrt{\int_0^\infty dR\, R^2[R\theta_{l_1l_2}(R)]^2/S_{l_1l_2}}$,
listed in Table~\ref{tab:sfpw}.
The SA radii of the dominant channel of the $0_2^+$
and $2_2^+$ states are 5.84 fm with $(l_1,l_2)=(0,0)$
and 7.38 fm with $(l_1,l_2)=(0,2)$, respectively.
Since the rms distance of the $^{8}$Be wave function 
is 5.32 fm, the $^8{\rm Be}+\alpha$ configuration
induces an acute-angled triangle geometry.

\section{Conclusion}
\label{conclusions.sec}

How similar is the structure of the $2_2^+$ state in the $^{12}$C
as compared to the Hoyle state?
We have made comprehensive investigations of 
the structure of $^{12}$C with a special emphasis on
the geometric configurations of $\alpha$ particles.
The $0^+$ and $2^+$ states of $^{12}$C are described
by a three-$\alpha$ cluster model with the orthogonality constraint.
Precise three-$\alpha$ wave functions are obtained by
using the correlated Gaussian expansion with the stochastic variational
method. We introduce a confining potential
to obtain a physical state, allowing us
to visualize the three-$\alpha$ configuration
by using square-integrable basis functions.

In comparison of the two-body density distributions of the $0_2^+$ and $2_2^+$ state,
the main three-$\alpha$ configurations are found to be the same;
the acute-angled triangle shape
coming from the $^{8}{\rm Be}(0^+)+\alpha$ component.
However, the magnitude is significantly smaller
for the $2_2^+$ state compared to the $0_2^+$ state.
We find that the $2^+_2$ state can be mainly excited
by the relative coordinate between $^{8}{\rm Be}$ and $\alpha$
consistently with the interpretation given in Refs~\cite{Uegaki77,Uegaki78,Uegaki79,Kanada07}.
The $^{8}$Be cluster in the $0_2^+$ state is hardly excited
because the excitation energy of the $^8$Be(2$^+$)
is higher than the energy difference of $2_2^+$ state from the Hoyle state.
Therefore, we conclude that the $2_2^+$ state is not an
ideal rigid Hoyle band but could be interpreted as
a partially rotational excited state of $0_2^+$.
We note, however,
that this does not contradict the $\alpha$-particle
mean-field picture for the $2_2^+$ state~\cite{Yamada05}.
It is interesting to study the $4^+_2$ state,
which is observed recently~\cite{Freer11} and
considered also as a candidate of the Hoyle band member.

\begin{acknowledgement}
This work was in part supported by JSPS KAKENHI Grants
Nos.\ 18K03635 and 22H01214.
We acknowledge the Collaborative Research Program 2022, 
Information Initiative Center, Hokkaido University.
\end{acknowledgement}

\end{document}